\documentclass[10pt]{article}

\usepackage[dvips]{graphicx}
\usepackage{amsmath}
\usepackage{float}

\newcommand{\He}{\rm{H}}

\newcommand{\rmd}{{\rm d}}
\newcommand{\rme}{{\rm e}}

\newcommand{\mbfj}{\mathbf{j}}

\newcommand{\mbfr}{\mathbf{r}}

\newcommand{\mbfE}{\mathbf{E}}

\newcommand{\mbfsigma}{\boldsymbol{\sigma}}

\newcommand{\Elf}{{\cal E}}
\newcommand{\ELF}{{\boldsymbol{\cal E}}}
\newcommand{\mgf}{{\cal B}}
\newcommand{\MGF}{{\boldsymbol{\cal B}}}

\newcommand{\AVP}{{\boldsymbol{\cal A}}}

\begin{document}

\title{A heuristic quantum theory of the integer quantum Hall effect}
\author{Tobias Kramer\\
Department of Physics\\
Harvard University\\
17 Oxford Street\\
Cambridge, MA 02138, USA.\\
Email: tobias.kramer@mytum.de}
\date{October 30, 2005}

\maketitle

\begin{abstract}
Contrary to common belief, the current emitted by a contact embedded in a two-dimensional electron gas (2DEG) is quantized in the presence of electric and magnetic fields. This observation suggests a simple, clearly defined model for the quantum current through a Hall device that does not invoke disorder or interactions as the cause of the integer quantum Hall effect (QHE), but is based on a proper quantization of the classical electron drift motion. The theory yields a quantitative description of the breakdown of the QHE at high current densities that is in agreement with experimental data. Furthermore, several of its key points are in line with recent findings of experiments that address the dependency of the QHE on the 2DEG bias voltage, results that are not easily explained within the framework of conventional QHE models.
\end{abstract}


\section{Open problems in the quantum Hall effect.}

Despite more than 25 years of effort to understand the nature of the quantum Hall effect (QHE), no comprehensive theoretical description emerged that is unanimously accepted. For instance, von~Klitzing et.~al recently remarked: \textit{``[The edge-channel model of B\"uttiker] is widely used in textbooks on the integer QHE. We don't want to discuss it here, since recent experimental observations of the current distribution point to a somewhat different microscopic picture of the QHE.''} (\cite{Klitzing2005a}, p.~41. Translation by the author, the experiment Klitzing refers to is \cite{Ahlswede2002a}). Unsolved problems remain, in particular the origin of the breakdown of the QHE at high current densities and thus large electric Hall fields. 

\noindent
To cite again von~Klitzing et~al., \textit{``Not understood is the exact mechanism, which leads to a breakdown of the QHE above a critical current.''} (\cite{Klitzing2005a}, p.~43). Experimentally, the transition from a quantized resistivity to the non-quantized (classical) result is well
documented and even described in an empirical theory, which couples the electric Hall field
with the width of the quantized Hall plateaus. Note that this discussions refers to the global breakdown of the quantized Hall effect, not to a local heating effect.

Remarkably, the prominent theories of the QHE essentially neglect the presence of the electric Hall field in their underpinnings. Traditionally, theories of conduction invoke impurities and scattering processes in order to derive a constant drift velocity in the presence of an accelerating electric field. The resulting constant drift velocity is then used to establish Ohm's law, which states a proportionality between the current $j$ and the electric field $j=\sigma\;\Elf$.

However, transport in crossed electric and magnetic fields provides an important exception from the rule that scattering is needed in order to suppress an acceleration in an electric field. 

This article describes in detail a theory of the QHE, which incorporates the electric field from the beginning.
The inclusion of the electric field is done in a non-perturbative way and thus is not based on linear response theories for the conductivity (i.e.\ the Kubo formula). Instead, in Sect.~\ref{sec:electric} and App.~\ref{sec:hallmodel}, I construct a quantum theory of the transport in crossed electric and magnetic fields.

It explains the breakdown of the QHE as a function of the electric Hall field by an exact scaling law, which is in precise agreement with the breakdown law empirically found from experimental observations \cite{Kawaji1993a,Kawaji1996a,Shimada1998a}.

Very recent experiments in two different materials (GaAs/AlGaAs hetero\-structures \cite{Ilani2004a} and Silicon-MOSFETs \cite{Cobden1999a}) over a wide magnetic field range suggest a simple empirical linear law for the location of the quantized resistivity plateaus as a function of the applied gate voltage. However, conventional theories provide no straightforward explanation;  the proposed models make complicated assumptions in order to explain a simple relationship.

In the present theory I address fluctuations in the particle number and in the chemical potential (Fermi energy), Sect.~\ref{sec:fluc}. A quantitative computation of the resistivity is linked to the density of states in App.~\ref{sec:gdos}.
\begin{figure}[t]
\begin{center}
\includegraphics[width=0.7\textwidth]{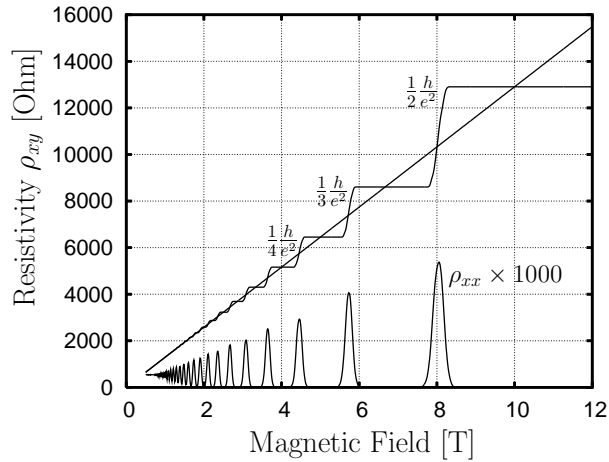}
\end{center}
\caption{
Classical Hall line (straight line) vs.~quantum Hall curve, calculated from Eq.~(\ref{eq:ConductivityTheory}).
The QHE leads to a quantized resistance $\rho_{xy}=\frac{1}{i}\frac{h}{e^2}$, $i=1,2,3,\ldots$.
Parameters (references for the values in brackets):
effective mass $m^*=0.1$,
mobility $\mu=17$~m$^{2}$V$^{-1}$s$^{-1}$,
effective $g$-factor $g^*=10$ \cite{EndNoteGFactor},
temperature $T=1$~K, 
current $j_x=1$~Am$^{-1}$,
average carrier density $N_{av}=2.4\times 10^{15}$~m$^{-2}$
(corresponding to a fixed Fermi energy of $E_F=11.6$~meV.)
}\label{fig:hallgraph}
\end{figure}
The QHE is thus put on a basis that does not rely on material properties like impurities. The only condition for the observation of the QHE (see Fig.~\ref{fig:hallgraph}) are low temperatures, clean samples, and the presence of a two-dimensional electron gas (2DEG). A schematic sketch of a MOSFET device, which may harbor a 2DEG, is given in Fig.~\ref{fig:hallsystem}.

In Sec.~\ref{sec:StraightLines} the present theory is shown to yield a natural and appealing quantitative solution for the linear variation of the plateau locations observed in \cite{Ilani2004a} and \cite{Cobden1999a}

\section{The role of the electric field in the quantum Hall effect.}\label{sec:electric}
\begin{figure}[t]
\begin{center}
\includegraphics[width=0.45\textwidth]{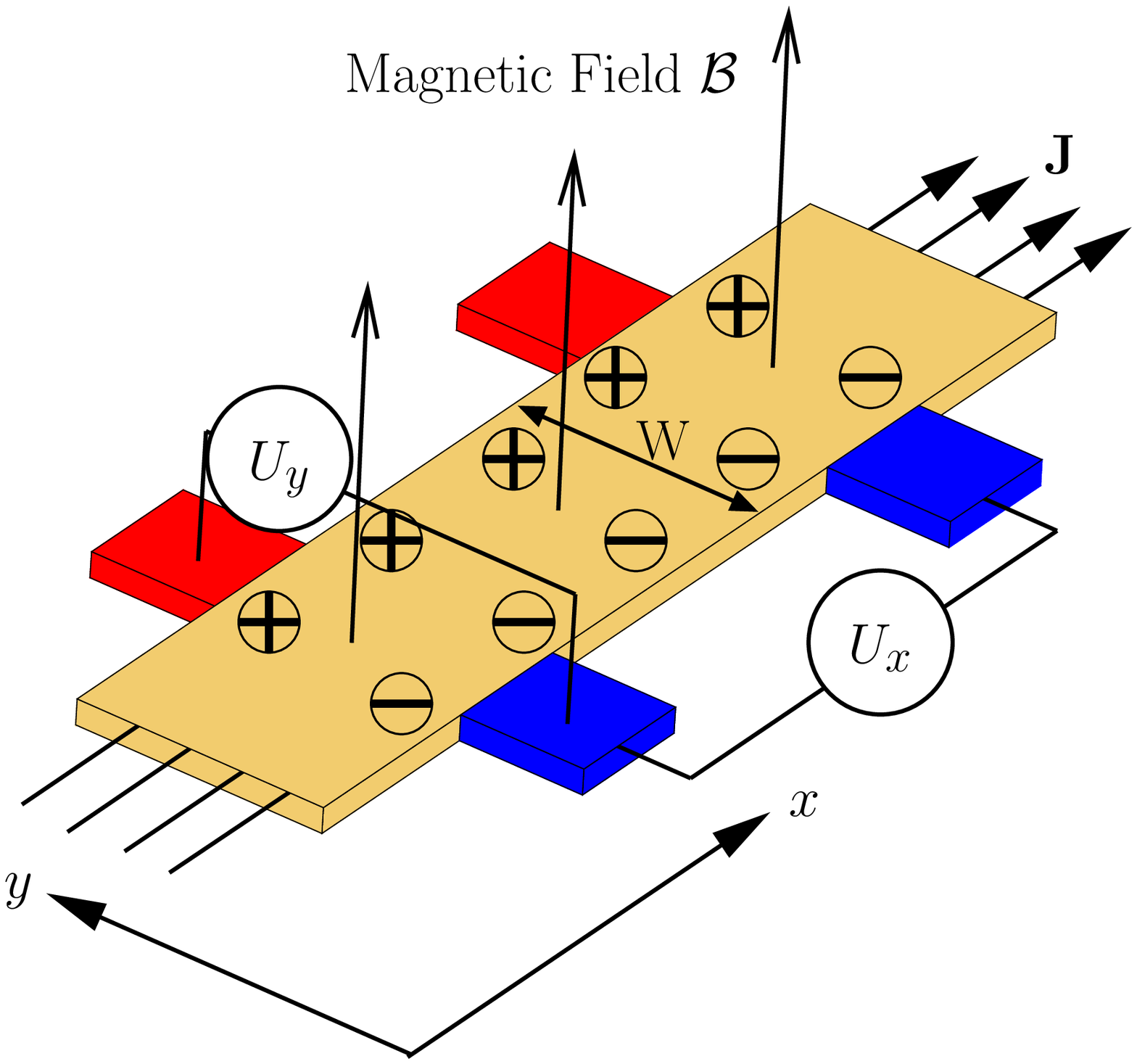}\hfill
\includegraphics[width=0.49\textwidth]{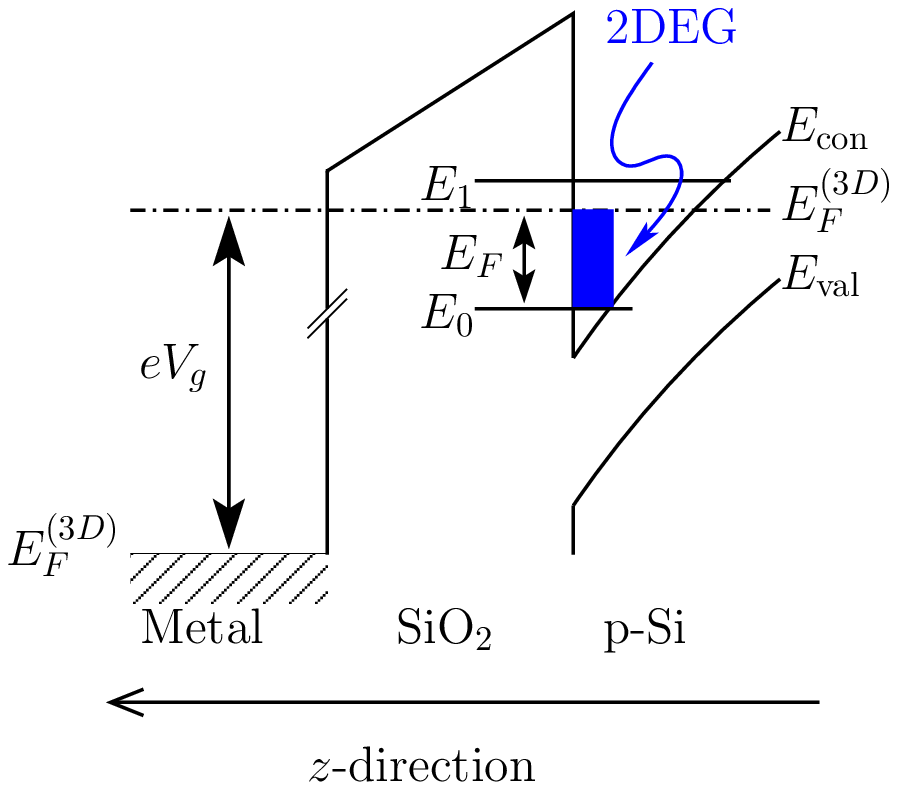}
\end{center}
\caption{
Left panel: Schematic view of a Hall bar. A current $\mathbf{J}$ is flowing through a two-dimensional electron gas (2DEG) in the $x-y$--plane, which is orientated perpendicular to an external magnetic field ${\cal B}$. The deflected electrons at the sample edges produce a Hall voltage $U_y$ over the sample width $W$, which is measured along with the longitudinal voltage drop $U_x$.
Right panel (adapted from \cite{Beenakker1991a}, Fig.~1): Schematic picture of a Metal-Oxide-Semiconductor (MOS) device. The two-dimensional electron gas (2DEG) at the interface between the oxide and the silicon is controlled by applying a gate voltage $V_g$. The gate voltage changes the Fermi energy $E_F^{(3D)}$ of the semiconductor, which in turn couples to the Fermi-energy $E_F$ of the 2DEG. If $E_F<(E_1-E_0)$ holds, the electrons only populate the ground state of the 1D quantum well in $z$-direction that has the eigenenergy $E_0$, which links both Fermi energies via $E_F=E_F^{(3D)}-E_0$.}\label{fig:hallsystem}
\end{figure}
In the classical Hall effect, the electric Hall field perpendicular to the current plays the crucial part in maintaining the electron drift motion, where the velocity is given by the ratio of the electric and magnetic fields (see App.~\ref{sec:HallClassic}).

\begin{figure}[t]
\begin{center}
\includegraphics[width=0.49\textwidth]{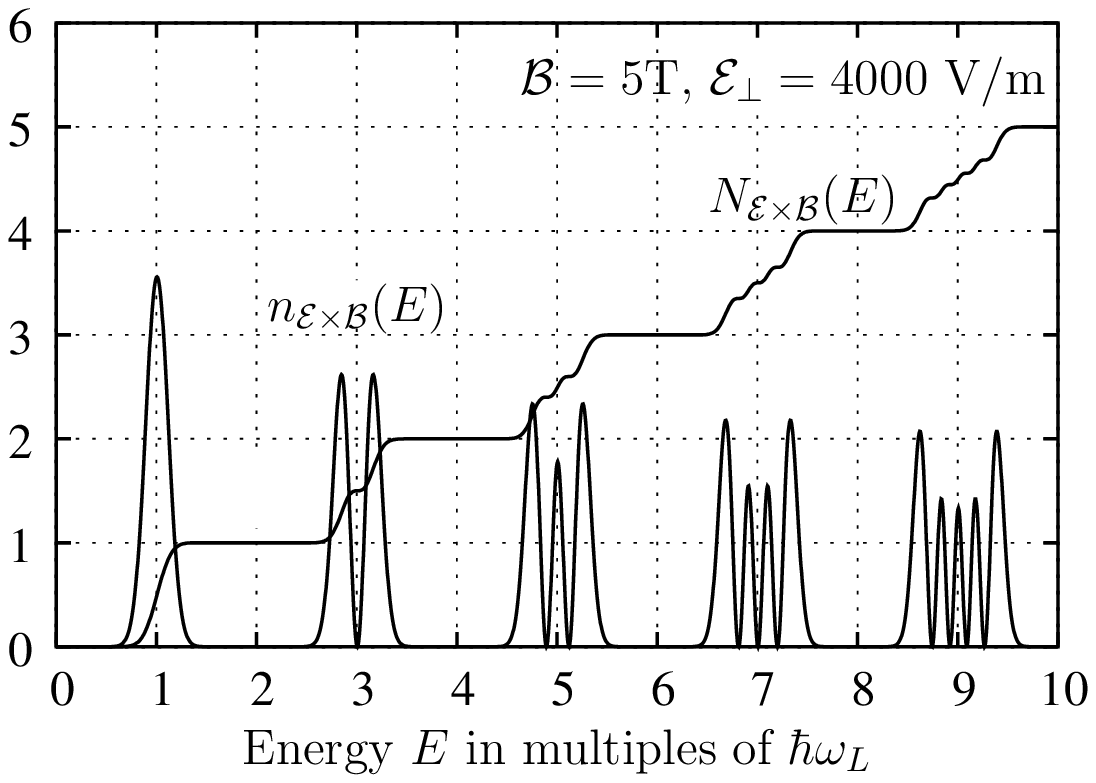}\hfill
\includegraphics[width=0.49\textwidth]{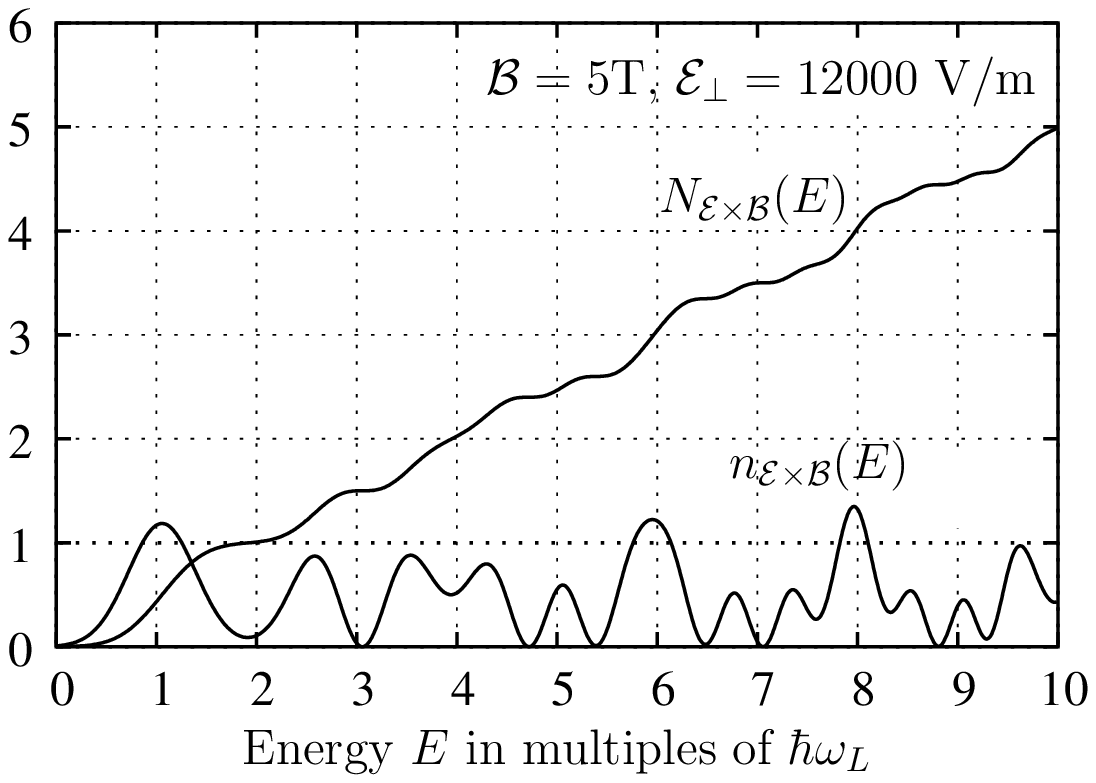}
\end{center}
\caption{Local density of states (LDOS) $n_{{\cal E}\times{\cal B}}(E)$ in crossed electric and magnetic fields, see eq.~(\ref{eq:DOSEB}) and \cite{Kramer2003b,Kramer2003a}. $N_{{\cal E}\times{\cal B}}(E)$ denotes the carrier density obtained from $N_{{\cal E}\times{\cal B}}(E)=\int_{-\infty}^{E}n_{{\cal E}\times{\cal B}}(E')\rmd E'$. Note the substructure within Landau levels and the broadening dependent on the electric field value. As pointed out in App.~\ref{sec:gdos}, the global density of states does not differ much from the LDOS.}\label{fig:doseb}
\end{figure}
In the classical (and quantum) Hall effect, the electric Hall field is the response of the system to an externally applied voltage. The resulting (longitudinal) current leads to a build up of charges near the edges of the sample, which in turn generate the electric Hall field that counterbalances the magnetic Lorentz force. Successive electrons propagate in the resulting steady-state crossed-fields configuration. While a theory of the QHE may include the (time-dependent) formation
of this steady-state, this process is not part of most theories.
The question, how the emittance of a localized contact is changed due to the presence of crossed electric and magnetic field is answered in Refs.~\cite{Kramer2003b,Kramer2003a,Kramer2003d}. The most remarkable feature is the complete suppression of a current between two Landau levels and also within a Landau level. These properties are reflected in a non-trivial form of the local density of states (see Fig.~\ref{fig:doseb}) and show the divergence of the quantum Hall effect from a classical electron drift picture: 
\begin{itemize}
\item For emission from a localized contact, the drift depends not only on the field ratio, but also on the kinetic energy of the electrons: for certain energy ranges, localized currents are formed with zero macroscopic flux and the electron propagation is blocked. This is in stark contrast to the classical case, where every electron can participate in the drift motion, independent of its initial (kinetic) energy.
\item Landau-levels are broadened by the electric field in a non-trivial way. Each Landau
level acquires a different substructure and width, dependent on the level number and the electric and magnetic field values (see Fig.~\ref{fig:doseb}).
\item The broadening follows a power law, which leads to a critical Hall field for the breakdown
\begin{equation}\label{eq:crit}
\Elf_{\rm crit}\propto B^{3/2}. 
\end{equation}
\item Higher Landau levels begin to overlap and therefore cannot sustain a quantized transport.
\end{itemize}
Experiments by Kawaji et al.\ \cite{Kawaji1993a,Kawaji1996a,Shimada1998a}, who studied the QHE and its breakdown as a function of the electric Hall field, are in precise agreement with the theoretical predictions. In fact, the power law (\ref{eq:crit}) is empirically deduced from the experimental data in \cite{Kawaji1993a}.
\begin{figure}[t]
\begin{center}
\includegraphics[width=0.45\textwidth]{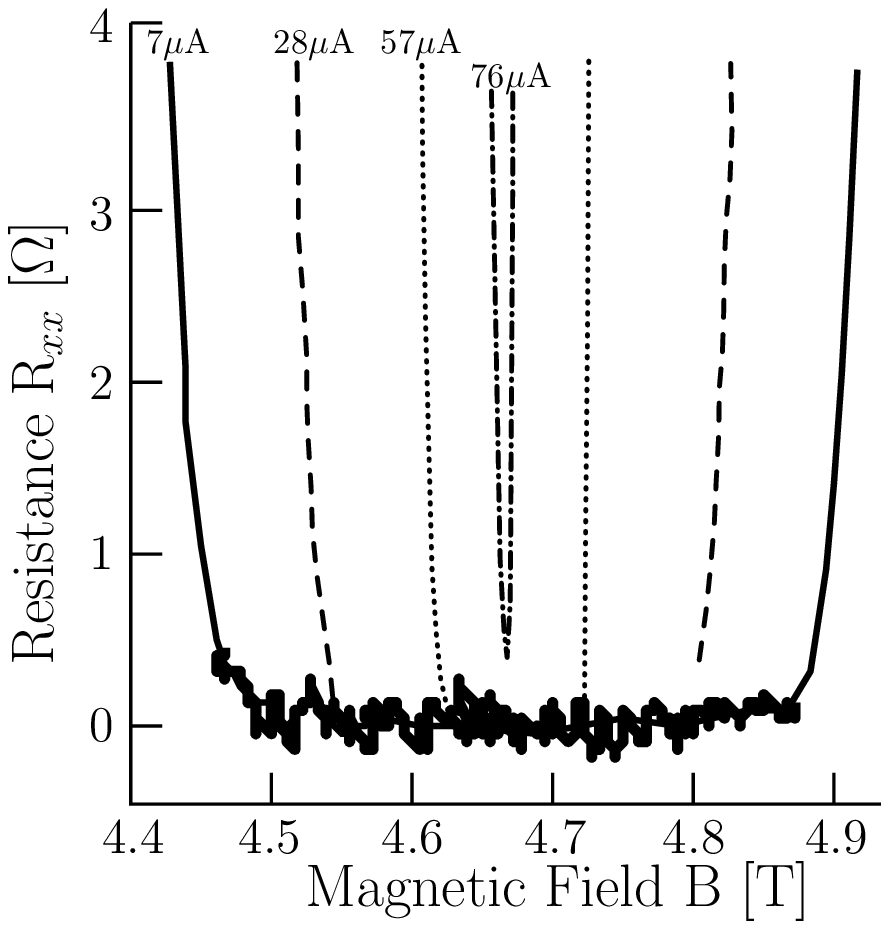}\hfill
\includegraphics[width=0.45\textwidth]{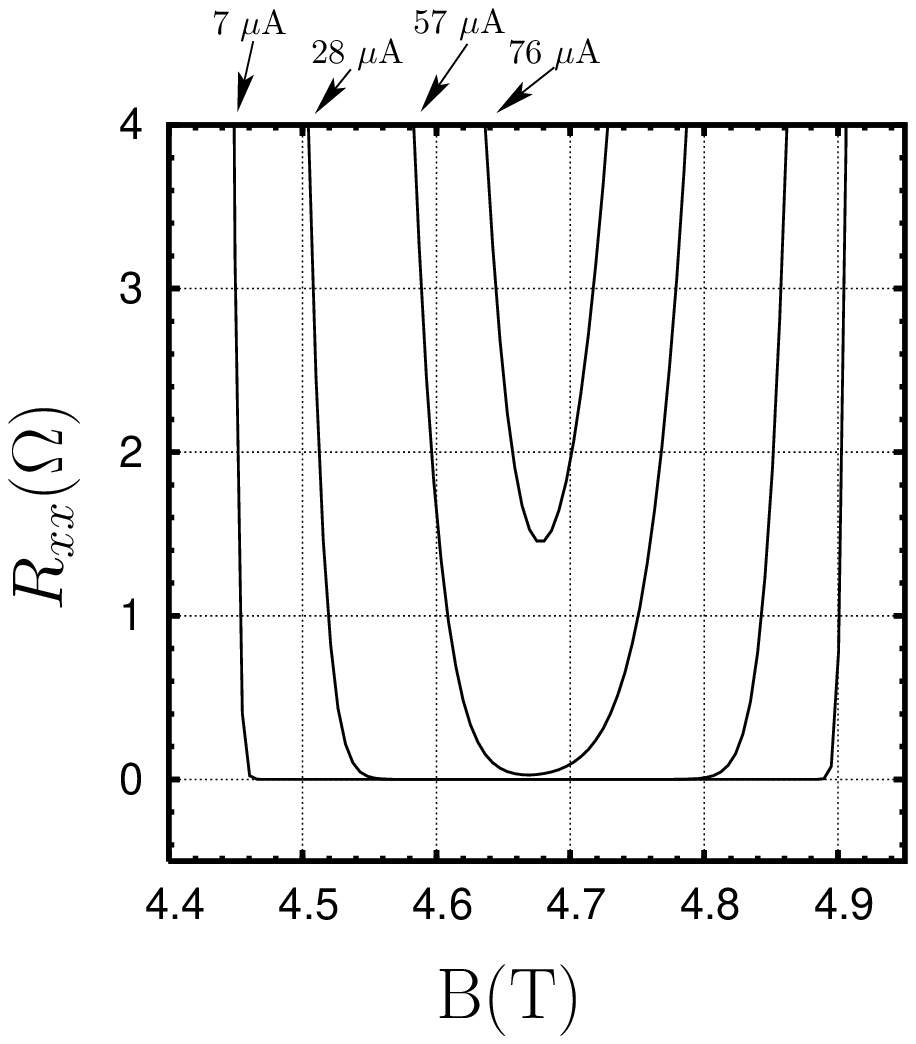}
\end{center}
\caption{Diagonal resistance $R_{xx}\propto\sigma_{xx}$ as a function of the magnetic field in the $i=4$ plateau for different currents and therefore electric Hall fields: $j_x=\sigma_{xy}(\mgf,\Elf)\,\Elf_y$. Left panel: A schematic representation of the experimental results obtained by Kawaji, published in \cite{Kawaji1996a}, Fig.~2. Right panel: theoretical prediction using the non-linear expression for the conductivity $\sigma_{xx}(E_F,\Elf_y,\mgf,T,\tau)$ derived in App.~\ref{sec:hallmodel} eq.~(\ref{eq:ConductivityTheory}), with the following parameters (references for the values in brackets): Effective mass $m^*=0.1$, scattering time $\tau=1\times 10^{-13}$~s, effective $g$-factor $g^*=12$ \cite{EndNoteGFactor}, temperature $T=1.2$~K, average number of particles $N_{av}=4.5\times 10^{15}$~m$^{-2}$ \cite{Kawaji1996a} (corresponding to a fixed Fermi energy of $E_F=10.7$~meV.) Due to the lack of more experimental data (i.e.\ over a wider magnetic field range), the parameters should be viewed as empirically derived. However, independent of the exact values, the observed power law for the critical Hall field (\ref{eq:crit}) is always reproduced by the theory.}\label{fig:breakdown}
\end{figure}
Also, Kawaji obtains different critical fields for different Landau levels, which is explained by the Landau-level dependent broadening in the theory \cite{Kramer2003a}.  The experimental findings can be explained within the heuristic theory of the Hall conductivity (see Eq.~(\ref{eq:ConductivityTheory})), which goes beyond linear response theories and their assumption of a linear relation between the conductivity and the current. Instead a non-linear relation
\begin{equation}
\mathbf{j}=\boldsymbol{\sigma}(\mgf,\Elf)\cdot\ELF
\end{equation}
is derived in App.~\ref{sec:hallmodel}. A comparison of the theory and experimental data is shown in Fig.~\ref{fig:breakdown}.

\section{Fluctuations of the particle number and the chemical potential.}\label{sec:fluc}
\begin{figure}[t]
\begin{center}
\includegraphics[width=0.45\textwidth]{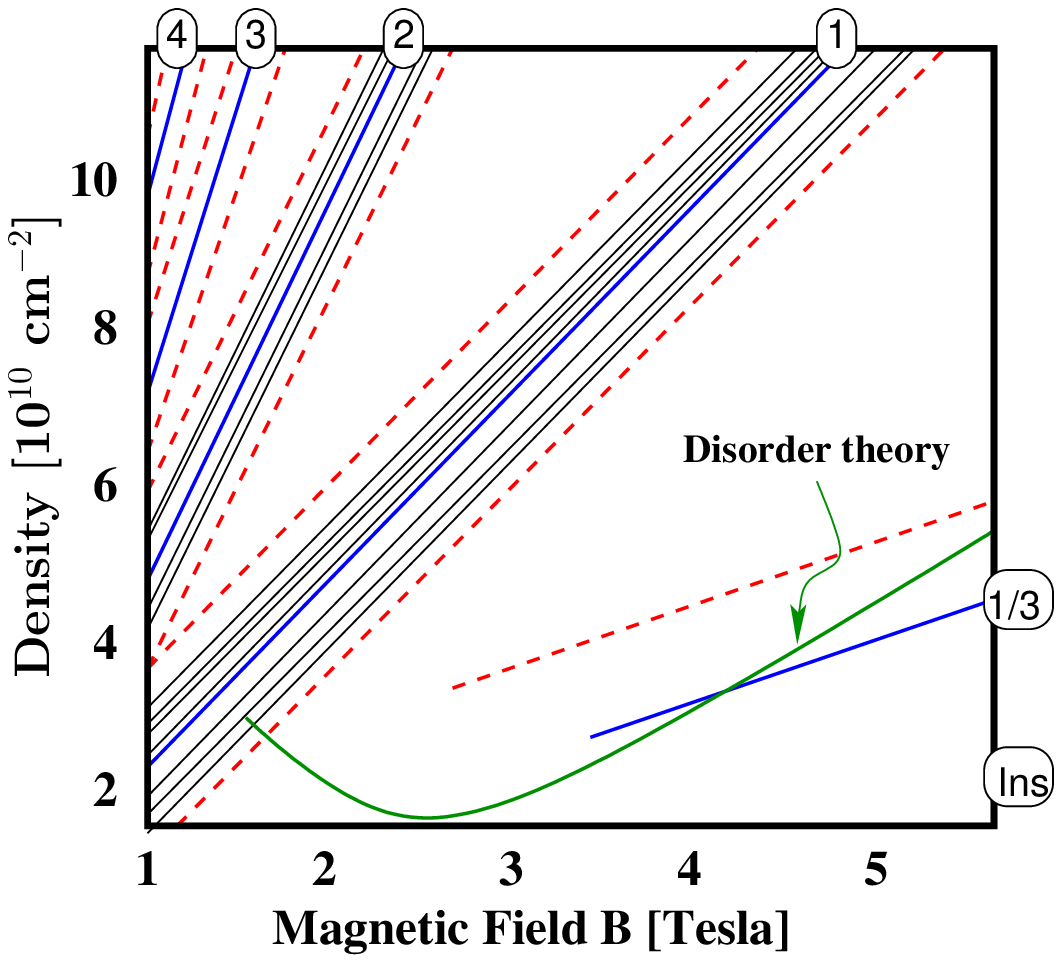}\hfill
\includegraphics[width=0.55\textwidth]{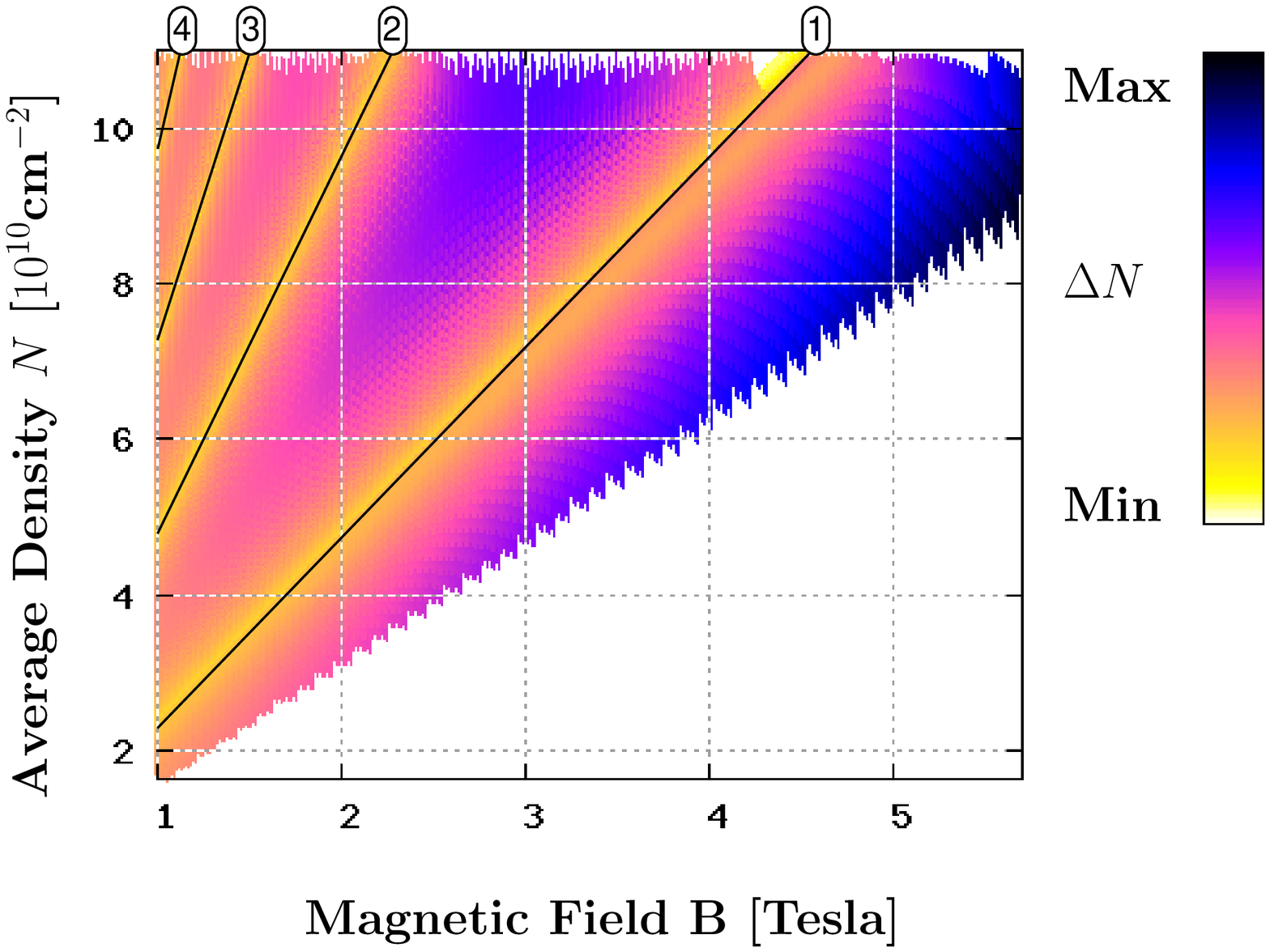}
\end{center}
\caption{Color plot of the fluctuation of the carrier density as a function of the magnetic field and an applied backgate voltage in a GaAs/AlGaAs heterostructure. Left panel: A schematic representation of the experimental data 
obtained by Ilani et al., published in \cite{Ilani2004a}, 
Fig.~1a. Right panel: theoretical prediction using eq.~(\ref{eq:ConductivityTheory}), with the following parameters (references for the values in brackets):
effective mass $m^*=0.07$ \cite{Mitin1999a},
mobility $\mu=50$~m$^{2}$V$^{-1}$s$^{-1}$ \cite{Ilani2004a},
effective $g$-factor $g^*=14.3$ \cite{EndNoteGFactor},
temperature $T=0.5$~K \cite{Ilani2004a}, 
current $j_x=0.1$~Am$^{-1}$ [assumed].
}\label{fig:ilani}
\end{figure}
%
\begin{figure}[t]
\begin{center}
\includegraphics[width=0.5\textwidth]{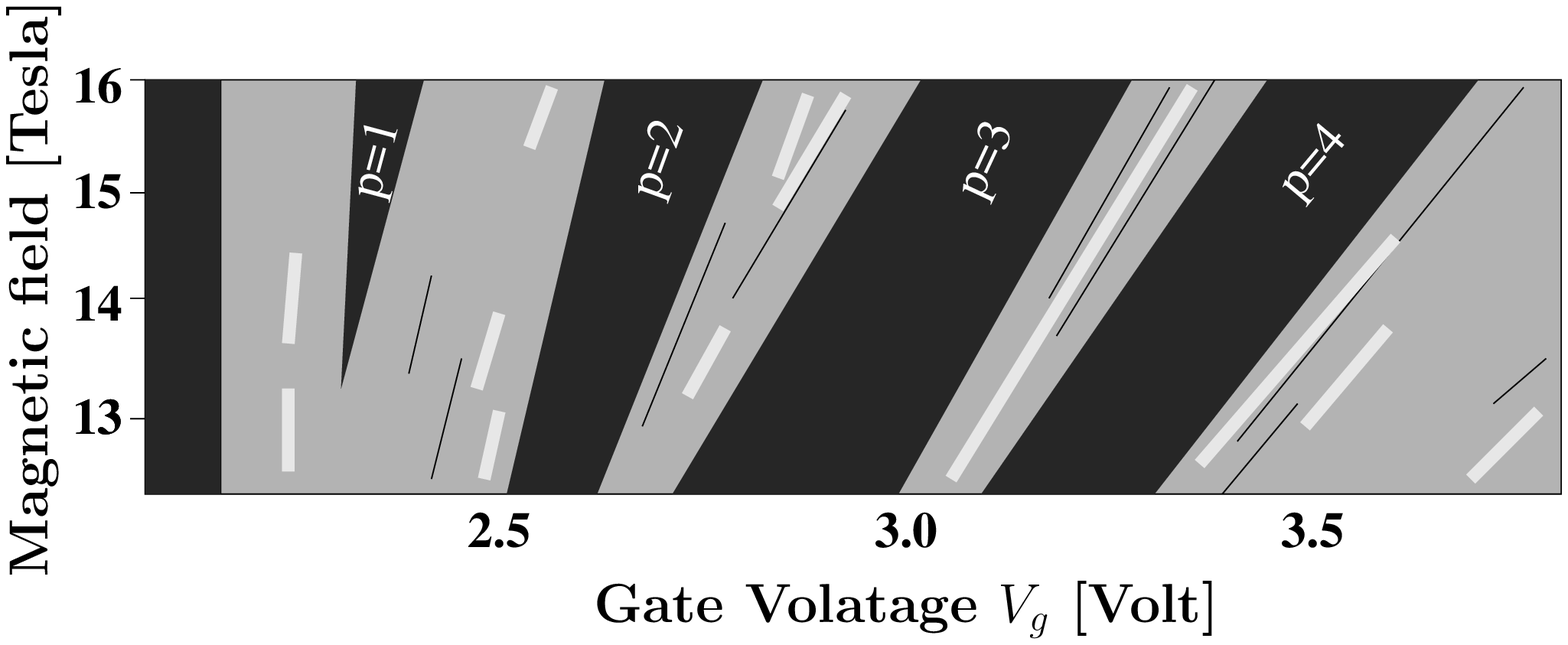}\hfill
\includegraphics[width=0.46\textwidth]{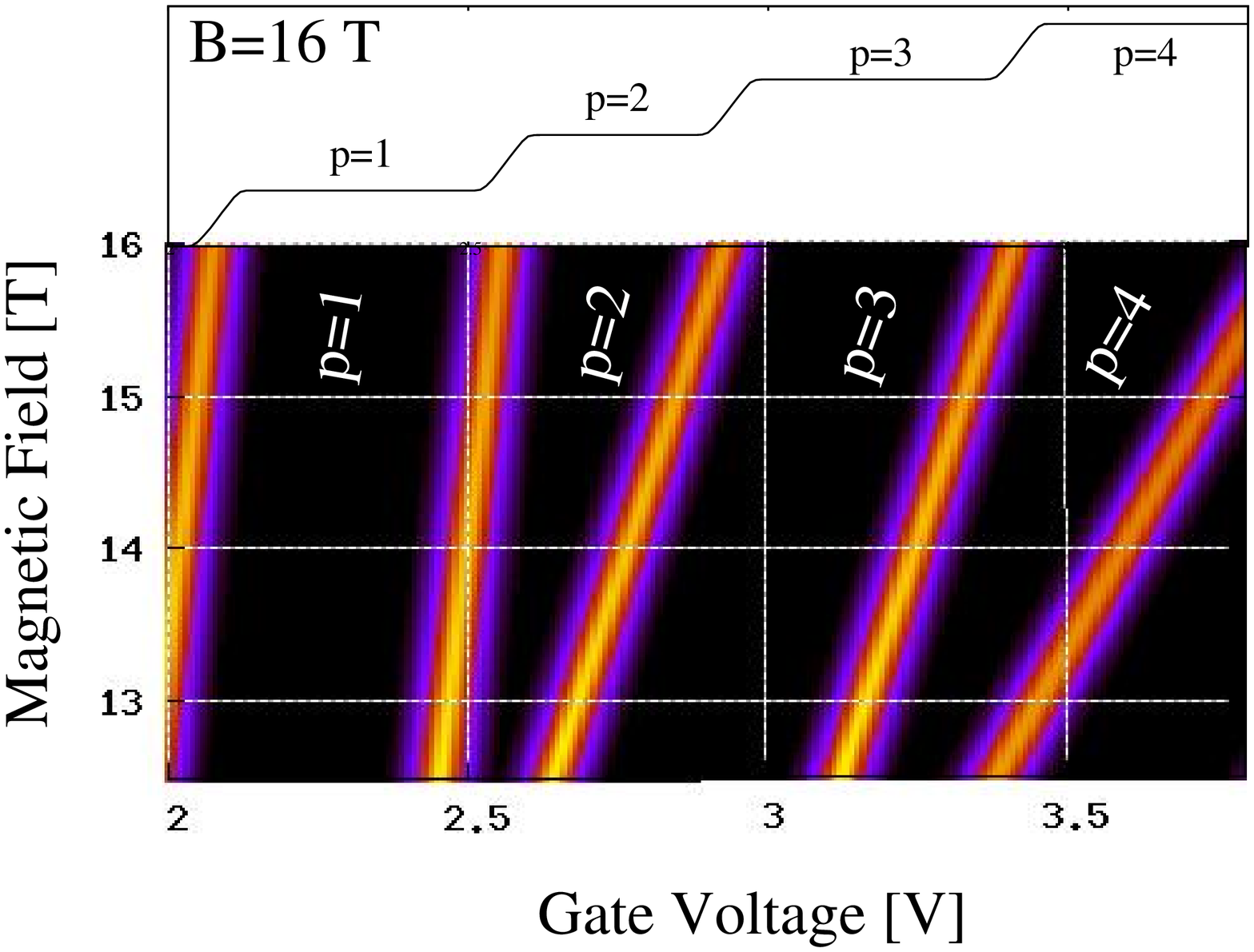}
\end{center}
\caption{Grayscale plot of the conductance $\sigma_{xx}$ as a function of the gate voltage $V_g$ and the magnetic field in a silicon MOSFET. Left panel: A schematic representation of the experimental data obtained by Cobden et al., published in \cite{Cobden1999a}, Fig.~2(a). Right panel: theoretical prediction using eq.~(\ref{eq:ConductivityTheory}), with the following parameters (references for the values in brackets):
transverse effective mass $m^*=0.19$ \cite{Mitin1999a},
mobility $\mu=0.19$~m$^{2}$V$^{-1}$s$^{-1}$ \cite{Cobden1999a},
effective $g$-factor $g^*=5$ \cite{EndNoteGFactor},
valley splitting in silicon $E_{\rm valley}=1.3$~meV \cite{Hirayama2003a},
temperature $T=1.0$~K \cite{Cobden1999a},
$C/e=8.6\times 10^{15}$~m$^{-2}$V$^{-1}$ \cite{Cobden1999a},
$V_{\text{off}}=2.3$~V \cite{Cobden1999a},
$j_y=0.1$~Am$^{-1}$ [assumed]. The location of the plateaus (enumerated by $p$) follows quantized slopes. In the transition region between two $p$'s, the theory shows less structure compared to the experimental result.
}\label{fig:cobden}
\end{figure}
Recent experiments probe the relationship between the location of the quantized plateaus and fluctuations in the number of carriers as a function of the magnetic field and an applied gate-voltage (see Fig.~\ref{fig:hallsystem}). The resulting pictures for two very different materials display a remarkably simple linear relation between the gate-voltage and the plateau location (left panels in Figs.~\ref{fig:ilani} and Fig.~\ref{fig:cobden}).

The challenge for a theoretical description is to connect the experimental parameter (the gate-voltage) with a theoretical quantity of the model. In Sec.~\ref{sec:StraightLines}, I propose and motivate a direct linear relation between the gate voltage $V_g$ and the Fermi energy $E_F$ in the system, which in turn determines the number of carriers $N$. The resulting calculations are in excellent agreement with the experimental results (right panels in Figs.~\ref{fig:ilani} and Fig.~\ref{fig:cobden})

\subsection{The connection between the Fermi energy and the density of states.}
The connection between the Fermi energy $E_F$ and the number of electrons $N$ is given via the local density of states (DOS). For the ground-state of a Fermi gas at zero temperature one obtains
\begin{equation}
N(E_F)=\int_{-\infty}^{E_F}\rmd E\,n(E).
\end{equation}
Early theories of the QHE were based on models of the DOS in a strong magnetic field. In the absence of external fields, the DOS of a free, non-interacting two-dimensional electron gas (2DEG) is independent of the energy of the state
\begin{equation}\label{eq:DOSfree}
n_{\text{free}}^{(2D)}(E)=\Theta(E)\frac{m}{2\pi\hbar^2},\quad
\Theta(E)=\left\{
\begin{array}{l}
0\quad E<0\\
1\quad E>0
\end{array}
\right.
\end{equation}
whereas for a purely magnetic field in two-dimensions, the DOS becomes a sum of delta-peaks
\begin{equation}\label{eq:DOSB}
n_{\mgf}^{(2D)}(E,\mgf)
=\frac{e \mgf }{2\pi\hbar}\sum_{k=0}^\infty
\delta\left(E-\hbar\omega_L[2k+1]\right),\quad \omega_L=\frac{eB}{2m},
\end{equation}
while the three-dimensional result for the DOS is available in closed form \cite{Dodonov1975a}.
In order to connect the number of carriers with a macroscopic current through the sample, one needs additionally the average velocity $v$ of particles in the system. This average velocity is zero in a purely magnetic field (since the particles undergo circular motion). Therefore a potential term is needed to construct a DOS which allows the particles to propagate through the system.

Possible potentials are provided by:
\begin{enumerate}
\item The edges of the sample, along which one can establish skipping orbits  \cite{Buettiker1988a}.
\item Randomly distributed scatterers, which produce a potential landscape $V_\text{LD}(\mbfr)$ that allows electrons to drift through the system.
The periodic lattice potential and uncorrelated disorder potentials are often assumed to disappear on the average: $\int\rmd\mbfr\, V_\text{LD}(\mbfr)=0$ \cite{Hajdu1994a,KramerB2003a}. 
Note that no long-range electric field is present in this treatment.
\item The electric Hall field.
\end{enumerate}
It is interesting to note that the electric Hall field was discarded as a possible candidate by \cite{Hajdu1994a,Thouless1998a,Yoshioka2002a}, since it was argued that the classical expression for the drift velocity is equal to the velocity given by the expectation value of $\mathbf{p}/m$ between eigenstates $|\psi\rangle$ of the Hamiltonian~(\ref{eq:HamiltonExB}). It was assumed that the only effect of the electric field is to cause this uniform drift. A modification of the density of states (DOS) caused by the electric field was not considered, because of the assumption of translational invariance of the system. If translational invariance was in place, the electric field could be eliminated by switching to a moving frame. Only impurities were considered as breaking the translational invariance, electric field effects on the transport were ruled out \cite{Prange1987a,Yoshioka2002a}. However, already the presence of stationary, i.e.\ non-moving contacts at the sample edges will break the translational symmetry (even without additionally introducing impurities). The modification of the DOS due to a electric field (see eq.~(\ref{eq:DOSEB}) and Fig.~\ref{fig:doseb}) yields besides a non-trivial broadening a substructure within each Landau level, which is discussed in \cite{Kramer2003a}.

\subsection{Experimental evidence for and against a fluctuation of the carrier density.}

In external fields, $N(E,\Elf,\mgf)$ is not constant, so a fixed Fermi energy $E_F$ implies a variable carrier density $N$ (or vice versa).
The question, whether $N={\rm const.}$ or $E_F={\rm const.}$ holds in a quantum Hall system, was investigated in different ways. A useful theoretical account was put forward by D.~Shoenberg, one of the pioneers in the field of magnetic oscillatory phenomena, who realized that the answer to this question is not easily given \cite{Shoenberg1984a}. The strong modulation of the DOS in the presence of a magnetic field gives rise to large fluctuations, either in the number of carriers or in the chemical potential, see Ref.~\cite{Kramer2003d}, Fig.~11. 

The disorder theory of the QHE tries to accommodate both, a fixed number of electrons and simultaneously a constant Fermi energy, by utilizing a reservoir of electrons from localized states. As Hajdu writes (\cite{Hajdu1994a}, p.~45):\\
\textit{
``However -- in contrast to free electrons -- by virtue of the disorder giving rise to level broadening, the electron concentration is a smooth function of the Fermi energy. In the case of free electrons both $\sigma_{yx}(E_F)=\sigma_{yx}^0(E_F)$ and $E_F(\nu)$ are step functions resulting in the classical straight line $\sigma_{yx}^0(\nu)\propto\nu$. Notice that $\sigma_{yx}(E_F)=\sigma_{yx}^0(E_F)$ together with $E_F(\nu)$ being a smooth function can be viewed as a condition for the QHE to occur.''}

Recent experiments \cite{Wei1997a,Weitz2000a} measure the variation of the electrostatic potential directly over a Hall bar. If the results are interpreted as a variations in the chemical potential, these findings contradict the condition for the occurrence of the QHE as cited above (disorder should buffer the oscillating Fermi energy and lead to a smooth function). On the other hand, fluctuations of the number of carriers $N(E_F)$ produce a charge which directly leads to a fluctuating electrostatic potential above the sample. In the disorder model, a systematic study of the carrier-density fluctuations for different gate-voltages results in complicated curves, which are \emph{not} observed experimentally (\cite{Ilani2004a}, Fig.~2a). However, I will point out a resolution within the heuristic model in Sec.~\ref{sec:StraightLines}.

In Ref.~\cite{Raymond1999a}, the authors measure the carrier concentration $N$ as a function of the magnetic field $\mgf$. Their results are in line with the model of a magnetic field dependent carrier concentration (i.e.\ $E_F={\rm const.}$ and \emph{not} $N={\rm const.}$).

Maybe the easiest way to develop a physical picture of the difference between the $N={\rm const.}$ and $E_F={\rm const.}$ models is based on the analogy with a capacitor. If the capacitor is an isolated system and not integrated into an electric circuit, charge conservation dictates $N={\rm const}$. However, as soon as the capacitor is attached to an electric circuit it operates in an $E_F={\rm const.}$ mode. This distinction has profound implications for the modeling of
devices via i.e.\ density functional theory (DFT) as pointed out in \cite{Lozovoi2001a}.

\section{The role of disorder and the location of the Hall plateaus.}\label{sec:StraightLines}

In the disorder model, localized, non-conducting states pin the Fermi energy between Landau levels. Along with the pinning mechanism, the plateau width (and location) also depends on the nature of the disorder. While 30 years ago only samples with impurities were available, the situation has changed dramatically: By using molecular beam epitaxy it is entirely possible to produce extremely clean samples, which nontheless still show the QHE \cite{Pfeiffer2003a}. Pfeiffer and West even suggest to artificially introduce disorder in the sample ``\textit{to begin a systematic study of the disorder problem}'' \cite{Pfeiffer2003a}.

The disorder problem is directly connected to the question where the Hall plateaus reside. Within the theory presented here, the location of the plateaus is given by the intersection points of the classical Hall line with the quantized Hall graph. At these points the classical and the quantum-mechanical expressions for the occupations coincide. The intersection points are readily calculated by equating both resistivities for the same Fermi energy $E_F$ 
\begin{eqnarray}\label{eq:rc}
R_{xy}^{cl}=\frac{\mgf}{e N_{\text{av}}} 
&\overset{!}{=}&
R_{xy}^{qm}=\frac{\mgf}{e \int_0^{E_F} n_{\ELF\times\MGF,\uparrow\downarrow}(E,\Elf,\mgf)\,\rmd E},
\\
N_{av}&=&\int_0^{E_F}\rmd E\,2\,n_{\text{free}}^{(2D)}(E)=\frac{m^*}{\pi\hbar^2} E_F,
\end{eqnarray}
where $n_{\text{free}}^{(2D)}(E)$ is given by (\ref{eq:DOSfree}) multiplied by two to account for the spin degeneracy and $n_{\ELF\times\MGF}(E,\Elf,\mgf)$ by eq.~(\ref{eq:DOSEB}) with the addition of a spin-splitting (see Sec.~5.5.2 in Ref.~\cite{Kramer2003d}). Note that the intersection point is not necessarily exactly in the middle of a plateau (see Fig.~\ref{fig:hallgraph}).

Recent experiments trace the evolution of the plateaus as a simultaneous function of the magnetic field $B$ and an applied gate-voltage $V_g$. Experiments have been performed using GaAs heterostructures  \cite{Ilani2004a} (see Fig.~\ref{fig:ilani}) as well as Silicon MOSFET devices \cite{Cobden1999a} (see Fig.~\ref{fig:cobden}). 

Both experiments confirm an extremely simple law for the plateau location in the $V_g$--$B$-plane: the plateaus follow straight lines, with a quantized slope. According to the authors, the simple disorder theories are not able to accommodate this linear dependency and much more complicated theories have to be invoked to recover a simple experimental result \cite{Ilani2004a}. Surprisingly, a straightforward solution is possible by using the $E_F={\rm const.}$ picture consistently and allowing for a variation of the particle number: If one assumes that the gate-voltage (minus a constant offset voltage) is strictly proportional to the Fermi-energy (even in the presence of a magnetic field)
\begin{equation}\label{eq:EfpropVg}
E_F=\alpha (V_g-V_o), 
\end{equation}
one obtains the observed quantized slopes as a function of the gate-voltage. 

The physical motivation between the assumption of a gate-voltage that directly adjusts the Fermi energy is shown in
Fig.~\ref{fig:hallsystem}. A Hall bar under current is part of an electric circuit, where external voltages fix the
Fermi level $E_F^{3D}$ of the device. The gate-voltage is translated into an effective Fermi energy $E_F$ of the two-dimensional subsystem under consideration (this is done by substracting the zero-point energy $E_0$ of the confining quantum well). At the same time an electric current is fed into the 2DEG subsystem via the contacts. In this sense, the
subsystem is treated as an open system (or in a broad sense ,,grand-canonical'' system, see also Sect.~{\ref{sec:fluc}}).

At the plateaus $R_{xy}^{qm}=\frac{h}{e^2\;i}, \quad i=1,2,3,\ldots$ holds and simultaneously one reaches the intersection point (\ref{eq:rc}) with the classical Hall line $R_{xy}^{qm}=R_{xy}^{cl}$. Therefore the magnetic field values at the intersection points with the quantized resistivity are given by
\begin{equation}\label{eq:Bi}
\frac{h}{e^2 i}=\frac{B}{e N_{av}} \Rightarrow B_i=\frac{h}{e\;i} N_{av}.
\end{equation}
An example for such an intersection point is $B_2=10$~T in Fig.~\ref{fig:hallgraph}. Now it is possible to derive how the magnetic field value of the intersection points changes as a function of the Fermi energy and therefore of the average particle number. Using Eq.~(\ref{eq:Bi}), I obtain for the slopes in GaAs/AlGaAs heterostructures
\begin{equation}
\frac{\partial B_i}{\partial N_{av}}=\frac{h}{e\;i},
\end{equation}
or expressing $\alpha$ in terms of the capacitance $C$ for a Si-MOSFET [$\alpha=C/(n_{\text{free}}^{(2D)}\,e)$]
\begin{equation}
\frac{e}{C}\frac{\partial B_i}{\partial V_g}=\frac{h}{e\;i}.
\end{equation}
These values reflect exactly the experimentally reported quantized slopes (\cite{Cobden1999a}, Eq.~(1), and \cite{Ilani2004a}, p.~329). 
Disorder was deliberately disregarded, although it may be important for the observed fine-structure in the experiments. A comparison of the theoretical prediction with experimental results is shown in Figs.~\ref{fig:ilani} and \ref{fig:cobden}. The excellent agreement supports the underlying model of a Fermi energy which is directly proportional to the applied gate voltage, while the actual number of particles may fluctuate about an average value.

\section{Summary and outlook.}
The heuristic theory presented in this article has features not contained in conventional theories of the QHE:
\begin{enumerate}
\item
It incorporates the electric field in the underlying density of states and yields the classical Hall effect in the limit of strong currents. It explains quantitatively the breakdown of the quantized Hall conductivity. Other theories do not consider the electric Hall field, and are thus unable to explain the (experimentally observed) dependency of the plateau width on the electric Hall field.
\item
The many-body aspect is taken into account by constructing a band model of the QHE, which is filled according to the density of states (DOS) in the presence of the external magnetic field and the electric Hall field. The DOS features gaps in the plateau regions.
\item
The current is calculated in a purely quantum-mechanical way, without using perturbative linear-response theory. The theory shows a sharp contrast between the classical propagation of electrons in crossed electric and magnetic fields emitted from a localized contact and their quantum-mechanical motion \cite{Kramer2003a,Kramer2003d}. 
\end{enumerate}
The electric field allows a current $J$ through the sample, which is calculated from the DOS under the assumption of an \emph{open system}:
\begin{itemize}
\item[4.]
In contrast to other theories of the QHE, this model allows for fluctuations of the number of carriers about an average value. The coupling between the Fermi energy of the two-dimensional electron gas and the device is provided by a gate voltage (see Fig.~\ref{fig:hallsystem}). The number of carriers is calculated as a function of the gate voltage (and therefore the Fermi energy). Note, that $N(E_F)$ will provide the plateaus, while the average drift velocity is constant (see also App.~\ref{sec:gdos}). As a result, $N(E_F)$ oscillates as a function of the magnetic field for fixed $E_F$. The gaps in the DOS in perpendicular electric and magnetic fields cause the observed conductivity quantization.
\end{itemize}
The difference of the actual occupation from the density of states is clearly formulated by Tanner \cite{Tanner1995a}, Sec.~2.1.2, who writes that the electron density is the product of two completely unrelated quantities, namely the density of states and the probability of occupation of a quantum state.
The probability distribution is a statistical function (i.e.\ given by the Fermi-Dirac distribution), whereas the DOS can be zero in some energy ranges. Moreover, \textit{``The Fermi energy $E_F$ can however still lie in that forbidden region because $E_F$ is simply defined as the energy, where the probability of occupation of a quantum state is $\frac{1}{2}$.''}

The model presented here does not depend on the presence of disorder for the formation of localized states. Instead, the crossed electric and magnetic fields block the propagation of the electrons in certain energy ranges. Since a real sample will never be completely clean, models for scattering processes should supplement the theory. A simple Drude-like approach is sketched in App.~\ref{sec:hallmodel}, but more sophisticated models are needed in order to explain the fine-structure which is visible in Figs.~\ref{fig:ilani} and \ref{fig:cobden}. One possible approach \cite{Donner2004a} is to model scatterers by properly regularized zero-range potentials \cite{Wodkiewicz1991a}.

One also should investigate whether the electric field is uniform and homogeneous across the sample. Experimentally, almost uniform Hall fields are observed for certain magnetic field ranges \cite{Ahlswede2002a}. Electrostatic models of the charge distribution in a Hall device point to the possibility of a sequence of metallic/non-metallic ``stripes'' in the sample \cite{Chklovskii1992a}. Such a Hall field distribution could be used as an input for the present model, see App.~\ref{sec:gdos} and Sec.~5.4 in Ref.~\cite{Kramer2003d}.

The last point concerns the fractional quantum Hall effect (FQHE). Surprisingly, crossed electric and magnetic field induce a substructure in a Landau-level which leads to plateaulike structures at several fractional and nearly fractional values of the conductivity quantum \cite{Kramer2003a}. Although their values match the observed FQHE fractions only partially, it is nevertheless remarkable that a non-interacting particle theory already generates a fractional pattern. The inclusion of interactions (possibly via electric Coulomb forces) in the present theory is surely desirable.

\subsection*{Acknowledgments}

I thank S.~Kawaji for pointing me to experimental data. Also I thank H.~Stumpf for the permission to
quote a passage from his book \cite{Stumpf1983a}.  I appreciate helpful discussions with C.~Bracher, M.~Kleber, and V.~Kanellopoulos. Financial support by the Deutsche Forschungsgemeinschaft (grants KL~315 and KR~2889 [Emmy Noether Program]) is gratefully acknowledged.

\appendix

\section{Brief review of the classical Hall effect.}\label{sec:HallClassic}

The propagation of electrons in crossed electric and magnetic fields has received a lot of attention in classical physics. The classical Hall effect is a standard method to determine the carrier density of a sample.  The basic idea in a classical Hall experiment is to detect the response of an electric current $\mathbf{J}$ to an applied magnetic field $\mgf$ \cite{Hall1879a}.

In the following the magnetic field $\MGF$ is aligned along the $z$ direction and the electronic current is confined in the $x$-$y$-plane, where an electric field $\Elf$ is present.\
\footnote{For the classical case, this assumption can be realized by using a thin strip of the material. In the quantum case, the effective two-dimensionality arises due to the formation of a quantum well, where only the lowest level is occupied. Thus one may speak of a two-dimensional system.}
Hamilton's function for crossed electric and magnetic fields reads:
\begin{equation}\label{eq:HamiltonExB}
H=\frac{{\left[\mathbf{p}-\frac{e}{c}\AVP(\mathbf{r})\right]}^2}{2m}
  -e\ELF\cdot\mathbf{r},\qquad
\AVP(\mathbf{r})=\frac{1}{2}(\MGF\times\mathbf{r})
\end{equation}
Without loss of generality one may align the electric field along the $y$-axis and obtain the following equations of motion ($\mathbf{r}(0)=\mathbf{o}$):
\begin{multline}
\mathbf{r}(t)=
\left[\begin{matrix}
x(t)\\
y(t)
\end{matrix}\right]
=
\frac{1}{\mgf^2 e}
\begin{pmatrix}
-\mgf p_y(0)         & -\Elf_y m + B p_x(0)\\
-\Elf_y m + B p_x(0) & \mgf p_y(0)
\end{pmatrix}
\cdot
\left[\begin{array}{c}
\cos(e\mgf t/m)\\
\sin(e\mgf t/m)
\end{array}\right]
\\
+
\left[
\begin{array}{c}
\frac{\Elf_y}{\mgf}t+\frac{p_y(0)}{\mgf e}\\
\frac{m\Elf_y}{\mgf^2 e}-\frac{p_x(0)}{\mgf e}
\end{array}
\right].
\end{multline}
The initial velocity vector is given by
\begin{equation}
\dot{\mathbf{r}}(t=0)=\frac{\mathbf{p}(0)}{m}
\end{equation}
The only time-dependency besides the cyclotron motion arises from the drift term $\Elf_y t/\mgf$ in $x$. The corresponding drift velocity (averaged over one period $T=2\pi m/(eB)$) reads
\begin{equation}\label{eq:DriftVelocity}
\mathbf{v_d}=\frac{1}{T}\int_{t}^{t+T}{\rm d}t'\,
\dot{\mathbf{r}}(t')=(\ELF\times\MGF)/\mgf^2.
\end{equation}
In a frame of reference that is moving with this velocity the otherwise trochoidal orbit becomes a circle:
\begin{equation}
{[\mathbf{r}(t)-\mathbf{r}_c(t)]}^2
=\frac{{[\Elf_y m-\mgf p_x(0)]}^2-{[\mgf p_y(0)]}^2}{\mgf^4 e^2},
\:
\mathbf{r}_c(t)
=
\left[
\begin{array}{c}
\frac{\Elf_y}{\mgf}t+\frac{p_y(0)}{\mgf e}\\
\frac{m\Elf_y}{\mgf^2 e}-\frac{p_x(0)}{\mgf e}
\end{array}
\right].
\end{equation}
Noting that the velocity in turn is related to the classical current $\mathbf{J}$ by the relation
\begin{equation}\label{eq:hc}
\mathbf{J}=N e \mathbf{v_d},
\end{equation}
where $N$ denotes the electron density and $e$ the electronic charge, one can extract the conductivity tensor $\mbfsigma$ from Ohm's law in the two-dimensional $(x,y)$-plane:
\begin{equation}\label{eq:ConductivityTensor}
\mathbf{J}=\mbfsigma\cdot\ELF
\quad\Rightarrow\quad
\mbfsigma=
\frac{N e}{\mgf}
\begin{pmatrix}
0 & -1\\
1 & 0
\end{pmatrix}
\end{equation}
This remarkable equation predicts a finite conductivity even in the absence of any scatterers, which are usually required in theories of conduction to establish an on average constant electron velocity. Also note that the drift velocity is independent of the kinetic energy of the electrons.

\section{Expression for the Hall resistance in the heuristic model.}\label{sec:hallmodel}

The formulation of the heuristic model is described in detail in \cite{Kramer2003a,Kramer2003d}. The basic quantity of interest is the DOS in crossed electric and magnetic fields, which is given by \cite{Kramer2003a}, eq.~(20):
\begin{equation}\label{eq:DOSEB}
n_{\ELF\times\MGF}(E) = \sum_{k=0}^\infty n_{k,\ELF\times\MGF}(E), \quad
n_{k,\ELF\times\MGF}(E) =
\frac{1}{2^{k+1} k! \pi^{3/2}l^2\Gamma} \, \rme^{-E_k^2/\Gamma^2} \, {\left[\He_k\left(E_k/\Gamma\right)\right]}^2,
\end{equation}
where $\He_k(x)$ denotes the Hermite polynomial. The level width parameter $\Gamma$ and the energies $E_k$ are given by
\begin{equation}
\Gamma = e \Elf_y \sqrt{\hbar/(e\mgf)},\quad E_k=E-\Gamma^2/(4\hbar\omega_L)-(2k+1)\hbar\omega_L.
\end{equation}
The quantum mechanical current for a system with Fermi-energy $E_F$ without scattering events and at zero temperature becomes \cite{Kramer2003d}, Sec.~5.5.1
\begin{equation}
j_x=\int_{-\infty}^{E_F}\frac{e\Elf}{\mgf}n_{\ELF\times\MGF}(E)\,\rmd E,
\end{equation}
from which one obtains by formal application of Ohm's law $\mbfj=\mbfsigma\cdot\mbfE$  the conductivities
\begin{eqnarray}\label{eq:cond0}
\sigma_{xy}(E_F,\Elf,\mgf)&=&
\frac{e}{\mgf }\int_{-\infty}^{E_F}\rmd E\;n_{\ELF\times\MGF}^{(2D)}(\mbfr;E),\\
\sigma_{xx}&=&0.
\end{eqnarray}
It is important to realize, that this expression couples the specific form of the density of states in crossed electromagnetic fields with the drift velocity $\Elf/\mgf$. It is not possible to introduce the occupation number $N_{\ELF\times\MGF}(E_F,\Elf,\mgf)=\int^{E_F}\rmd E\,n_{\ELF\times\MGF}(E)$ as a field independent classical parameter like in the classical Hall effect (\ref{eq:hc}). Also the conductivities include the full dependence on the electric fields, which is in general non-linear. This is in sharp contrast to approximative and perturbative methods like linear-response theories.

For a more realistic systems (i.e.\ at a finite temperature and with some scattering events), one has to supplement Eq.~(\ref{eq:cond0}) by a temperature $T$ and a scattering model. Since the temperature only affects the probability of occupying a state,
it is readily included by using the Fermi-Dirac distribution:
\begin{equation}\label{eq:FD}
f_{FD}(E,E_F,T)={\left[\rme^{(E-E_F)/(k_B T)}+1\right]}^{-1}.
\end{equation}
For the scattering, I assume the very simplistic Drude model, which employs a scattering time $\tau$ \cite{Ashcroft1976a}. Since scattering requires the presence of occupied and empty states, the Pauli principle restricts scattering processes to a small range around the Fermi-energy. This is reflected by the derivative of the Fermi-Dirac distribution \cite{Wei1985a} in Eq.~(\ref{eq:FDD}). Also, I assume that the scattering is proportional to the DOS via a constant $D$. Including both contributions leads to the following non-linear expressions for the conductivities:
\begin{eqnarray}\label{eq:ConductivityTheory}
\sigma_{xy}(E_F,\Elf_y,\mgf,T,\tau)
&=&
\frac{e}{\mgf}\int_{-\infty}^{\infty}\rmd E\;
n_{\ELF\times\MGF,{\uparrow,\downarrow,{\rm valley}}}^{(2D)}(E)
\frac{f_{FD}(E,E_F,T)}{1+{[2\omega_L\tau]}^{2}},\\\label{eq:FDD}
\sigma_{xx}(E_F,\Elf_y,\mgf,T,\tau)&=&\nonumber\\
&&\hspace{-18ex}\int_{-\infty}^{\infty}\rmd E\;
n_{\ELF\times\MGF,{\uparrow,\downarrow,{\rm valley}}}^{(2D)}(E)
\left(-\frac{\partial f_{FD}(E,E_F,T)}{\partial E}\right)\;
\frac{2\omega_L\tau D}{1+{[2\omega_L\tau]}^2}.
\end{eqnarray}
Finally, Hall conductivity and resistance are interrelated via:
\begin{eqnarray}
\rho_{xy}(E_F,\Elf_y,\mgf,T,\tau)&=& \frac{\sigma_{yx}}{\sigma_{xx}^2+\sigma_{xy}^2},\\
\rho_{xx}(E_F,\Elf_y,\mgf,T,\tau)&=& \frac{\sigma_{xx}}{\sigma_{xx}^2+\sigma_{xy}^2}.
\end{eqnarray}
In order to connect the model with material parameters, values for both the effective mass $m^*$ and the effective $g$-factor $g^*$ in a 2DEG are needed. Some materials exhibit additional degrees of degeneracy, i.e.\ silicon shows a valley splitting (see also Sec.~5.5.2 in \cite{Kramer2003d}). 

The last important point is the determination of the electric Hall field. If the experiment is conducted under constant current conditions, the implicit equation
\begin{equation}
\Elf_y=\rho_{xy}(E_F,\Elf_y,\mgf,T,\tau)\;j_x
\end{equation}
has to be solved for given $E_F$, $\mgf$, $T$, $\tau$, and $j_x$.

For the present article, I assume a homogeneous Hall field throughout the sample, which is coupled to the model of a position dependent Fermi energy \cite{Halperin1986a}. Experimental findings indicate, that this model is probably not generally valid, but has to be replaced by a spatially varying Hall field (and Fermi energy) \cite{Ahlswede2002a} that minimizes the electrostatic energy of the system. The current through a macroscopic device is given by integrating the current density over the device width $W$ (\cite{Kramer2003d}, Sec.~5.4):
\begin{equation}
I_x = \int_0^W\rmd y\;j_x(x,y) = \int_0^W \rmd y\; \sigma_{xy}(E_F,\Elf_y,\mgf,T,\tau)\;\Elf_y(y) \;.
\end{equation}
The implications of the global (sample-wide) averaging process are discussed in the next section.

\section{From a local DOS to a global DOS in the presence of an electric field.}\label{sec:gdos}

One possibility to obtain the number of carriers in the sample from a known local DOS is to perform a spatial integration over the sample dimensions $L\times W$ \cite{Lannoo1991a}, p.~36:
\begin{equation}\label{eq:nav_wrong}
N_{\text{global}}^{\text{space}}=
\int_0^{E_F} \rmd E 
\left[
\int_{0}^{W} \rmd y 
\int_{0}^{L} \rmd x \; n(x,y;E)
\right].
\end{equation}
One may argue that this process is similar to an averaging process over the energy, since $y\Elf_y=\Delta E$. If $W\Elf_y\gg 2 \hbar\omega_L$, the Landau-quantization is effectively washed out and no macroscopic quantization of the resistivity will remain. 

Since the above argument applies to any model of the QHE, which is based on a Landau quantization, one has to address this concern carefully. A detailed analysis is given by Stumpf \cite{Stumpf1983a}, p.~638f, who examines the expression for the current
\begin{equation}
J=N\int\,e\,\mathbf{v}(\mathbf{K})\,\tilde{f}(\mathbf{K},t)\rmd^3K,
\end{equation}
where $\mathbf{v}$ denotes the velocity, $\mathbf{K}$ the wave vector, and
\begin{equation}\label{eq:Stumpf1}
\int\tilde{f}(\mathbf{K},t)\rmd^3K=1
\end{equation}
the distribution function. While for a classical system, $\mathbf{v}$, $f$ and $N$ are decoupled (see i.e.\ Eq.~(\ref{eq:hc})), a quantum mechanical theory leads to a coupling. Stumpf writes:\\
\textit{Now in a stationary state of a system the number $N$ is not independent of the processes occurring outside of the conduction band. Rather its mean value $\bar{N}$ is a stationary solution of rate equations derived for processes going on at impurities. Such solutions obviously depend on temperature, external fields, etc. Thus we have
\begin{equation}
\bar{N}:=N(T,\ELF,\ldots)
\end{equation}
and the number of conduction band electrons therefore depends on the physical state of the ``underground'' processes. Hence it is obvious that we have to replace [(\ref{eq:Stumpf1})] by the condition
\begin{equation}\label{eq:Stumpf2}
\int f(\mathbf{K},t)\rmd^3K=N(T,\ELF,\ldots).
\end{equation}
Thus the dependence of the current on the external forces does not only come about by the conduction band kinetics themselves but also by the normalization.}
 
Additionally to the influence of the external fields on the density of states, one has to establish an spatial averaging procedure over the sample in order to obtain a macroscopic current \cite{Stumpf1983a}:\\
\textit{A further generalization of (\ref{eq:Stumpf2}) is necessary if for a homogeneous medium the external forces depend on $\mathbf{r}$. In this case the statistical unit cells of the ensemble (microblocks, mosaicblocks) have to be correlated and $f$ depends on the number $i$ of the units cell or in macroscopic coarse-graining on $\mathbf{r}$ i.e., 
$f\equiv f(\mathbf{K},\mathbf{r},t)$. In this case also $\bar{N}$ depends on $i$ or $\mathbf{r}$, resp., i.e., $\bar{N}=N(\mathbf{r},T,\ELF,\ldots)$ and (\ref{eq:Stumpf2}) can be written as
\begin{equation}
\int f(\mathbf{K},\mathbf{r},t)\rmd^3 K=N(\mathbf{r},T,\ELF,\ldots)\,.
\end{equation}
}
I implement Stumpf's reasoning by discarding (\ref{eq:nav_wrong}) in favor of
\begin{equation}\label{eq:nav}
N_{\text{global}}^{\text{energy}}=
\int_{0}^{L} \rmd x 
\int_{0}^{W} \rmd y 
\left[
\int_0^{E_F(y)} \rmd E \; n_{\MGF\times\ELF(y)}(x,y;E).
\right].
\end{equation}
In a multi-electron system, the Pauli-principle guides us in occupying the available energy-levels. This integration should be performed first, before averaging over different positions. This procedure allows the introduction of the position-dependent Fermi energy via the following coarse-graining procedure (see also \cite{Kramer2003a}, Sec.~3):
\begin{equation}
\label{eq:tilt}
E_F(\mathbf{r})= E_F(\mathbf{o}) - q \mathbf{r}\cdot\ELF
\end{equation}
The crossed electromagnetic field configuration produces a ``tilted'' Fermi level \cite{Halperin1986a}:
an equilibrium in the sense of a overall constant Fermi energy cannot be reached, since a (ballistic) current flow along the Hall field is not possible due to the current deflection in the crossed field configuration.

\pagebreak

\bibliographystyle{unsrt}

\begin{thebibliography}{10}
\bibitem{EndNoteGFactor}
The effective $g$-factor is empirically deduced. I assume a constant $g^{*}$,
  which explains the observed structures very well. Thus a dependency of
  $g^{*}$ on the magnetic field is not needed here.
\end{thebibliography}

\providecommand{\url}[1]{#1}

\end{document}